\documentclass{iau307}
\usepackage{graphicx}
\usepackage{natbib}
\usepackage{url}
\usepackage{dtklogos}
\bibpunct{(}{)}{;}{a}{}{,}

\title[Asteroseismology \& spectropolarimetry: probing the dynamics of massive stars]
{Asteroseismology and spectropolarimetry: opening new windows on the internal dynamics of massive stars}

\author[S. Mathis \& C. Neiner]   
{S. Mathis$^{1,2}$
 \and C. Neiner$^2$}

\affiliation{$^1$Laboratoire AIM Paris-Saclay, CEA/DSM - CNRS - Universit\'e Paris Diderot, IRFU/SAp Centre de Saclay, F-91191 Gif-sur-Yvette Cedex, France\\ email: {\tt stephane.mathis@cea.fr} \\[\affilskip]
$^2$LESIA, Observatoire de Paris, CNRS UMR 8109, UPMC, Univ. Paris-Diderot,\\ 5 place Jules Janssen, 92195 Meudon, France}

\pubyear{2014}
\volume{307} 
\pagerange{}
\jname{New windows on massive stars: asteroseismology, interferometry, and spectropolarimetry}
\editors{G. Meynet, C. Georgy, J.H. Groh \\ Ph. Stee, eds.}

\begin{document}

\maketitle

\begin{abstract}
In this article, we show how asteroseismology and spectropolarimetry allow to probe dynamical processes in massive star interiors. First, we give a summary of the state-of-the-art. Second, we recall the MHD mechanisms that take place in massive stars. Next, we show how asteroseimology gives strong constraints on the internal mixing and transport of angular momentum while spectropolarimetry allows to unravel the role played by magnetic fields.

\end{abstract}

\section{New probes of the dynamics of massive stars}

Massive stars are dynamical objects: they are rotating, turbulent, pulsating, magnetic for 7\% of them, and they strongly impact their environment, because of their winds (ud-Doula, this volume) and tides (Auclair-Desrotour et al., this volume; Langer, this volume). These dynamical processes drastically affect their evolution and those of their galactic environment \citep[][]{Maeder2009}. Therefore, they must be modeled with a high degree of realism \citep[][]{Mathis2013} taking into account the best available observational constraints. In this context, our knowledge of stellar dynamics has undergone a revolution thanks to the seismology of the Sun and of stars \citep[e.g. SOHO, MOST, CoRoT, and {\it Kepler};][]{Aertsetal2010} and ground-based high-resolution spectropolarimetry that characterizes the magnetic field of stars at their surface \citep[e.g. ESPaDOnS/CFHT, Narval/TBL, HARPSpol/ESO;][]{DonatiLandstreet2009}. On one hand, {\it Kepler} demonstrated that very powerful mechanisms are acting to extract angular momentum from the radiative cores of sub-giant and red giant stars during their evolution \citep[][]{Becketal2012,Deheuvelsetal2012,Mosseretal2012,Deheuvelsetal2014}. This result creates a bridge with the case of main-sequence stars where such mechanisms are also needed to understand the flat rotation profile of the solar radiative core until $0.2R_{\odot}$ discovered by helioseismology \citep[][]{Garciaetal2007} and the weak differential rotation recently revealed by asteroseismology in some intermediate-mass and massive stars \citep[][for KIC$\,11145123$ and KIC$\,10526294$ respectively]{Kurtzetal2014,Papicsetal2014}. On the other hand, the MiMeS spectropolarimetric survey \citep[][]{Wadeetal2013} has demonstrated that only 7\% of massive stars host a large-scale magnetic field. Because of their stability along time, their simple geometry, which is often an oblique dipole, and their presence on the PMS with the same proportion of occurrence \citep{Alecianetal2013}, we concluded that they must be of fossil origin. These groundbreaking discoveries thus open a golden age for the study of dynamical processes driving the evolution of stars. In this context, the selection of new space and ground-based facilities, among which K2 (NASA), TESS (2017/NASA), SPIRou (2017/CFHT), and PLATO (2024/ESA), calls for an urgent progress in these fields of investigation and the corresponding theoretical modeling effort along this roadmap.

\section{Dynamical processes in massive stars}

On one hand, the core of massive stars is convective because of the strong exo-thermic behavior of the CNO cycle. Because of their inertia, turbulent coherent convective structures, the plumes, penetrate into the surrounding radiative envelope. This is the so-called overshoot. Then, mixing occurs at the convection-radiation boundary and internal gravity waves (hereafter IGWs) are stochastically excited \citep[e.g.][Meakin, this volume; Arnett, this volume; Mathis \& Neiner, this volume]{Browningetal2004}. Moreover, magnetic fields are generated in the core by a dynamo action because of sustained helical flows and differential rotation \citep{BBT2005}. If the dynamo field interacts with a surrounding fossil field, a strong-dynamo regime is obtained and its amplitude is increased \citep{Fetal2009}. Finally, it tends to inhibit the overshoot.  

On the other hand, four major transport mechanisms in stellar radiation zones must be studied. First, large-scale meridional flows, which are driven by applied torques and internal stresses, must be properly modelled \citep[e.g.][]{Zahn1992,MZ2004}. Then, various
hydrodynamical instabilities that generate turbulence in stably stratified layers must be identified and
characterized \citep[e.g.][]{Zahn1983,Metal2004}. Next, the effects of fossil magnetic fields and of their instabilities should be examined \citep[][Gellert, this volume]{MZ2005,Zetal2007}. Finally, IGWs must be taken into account. They are excited at the interfaces between convective and radiative regions by penetrative convection and by the $\kappa$-mechanism if stars are in the instability strip. They propagate in the radiation zone until they deposit/extract angular momentum where they are damped at their co-rotation layers \citep[e.g.][]{MdB2012,AMD2013}. In this context, dynamical stellar evolution codes that include differential rotation, meridional flows, shear-induced turbulence, and IGWs in a coherent way have been developed for fifteen years \citep{MM2000,TC2005,Detal2009,Metal2013}.

\section{Asteroseismology: new windows on internal mixing and transport of angular momentum in massive stars}

Asteroseismology is currently revolutionizing our vision of stars. It is now able, thanks to high-precision space photometry and ground-based spectroscopy, to provide stratified information on the structure, chemical composition, and dynamics of stellar interiors, which would be unavailable otherwise. 

\subsection{Internal mixing in massive stars: the case of HD\,181231 and HD\,175869}

\begin{figure}[h!]
\centering
\includegraphics[width=0.42\textwidth]{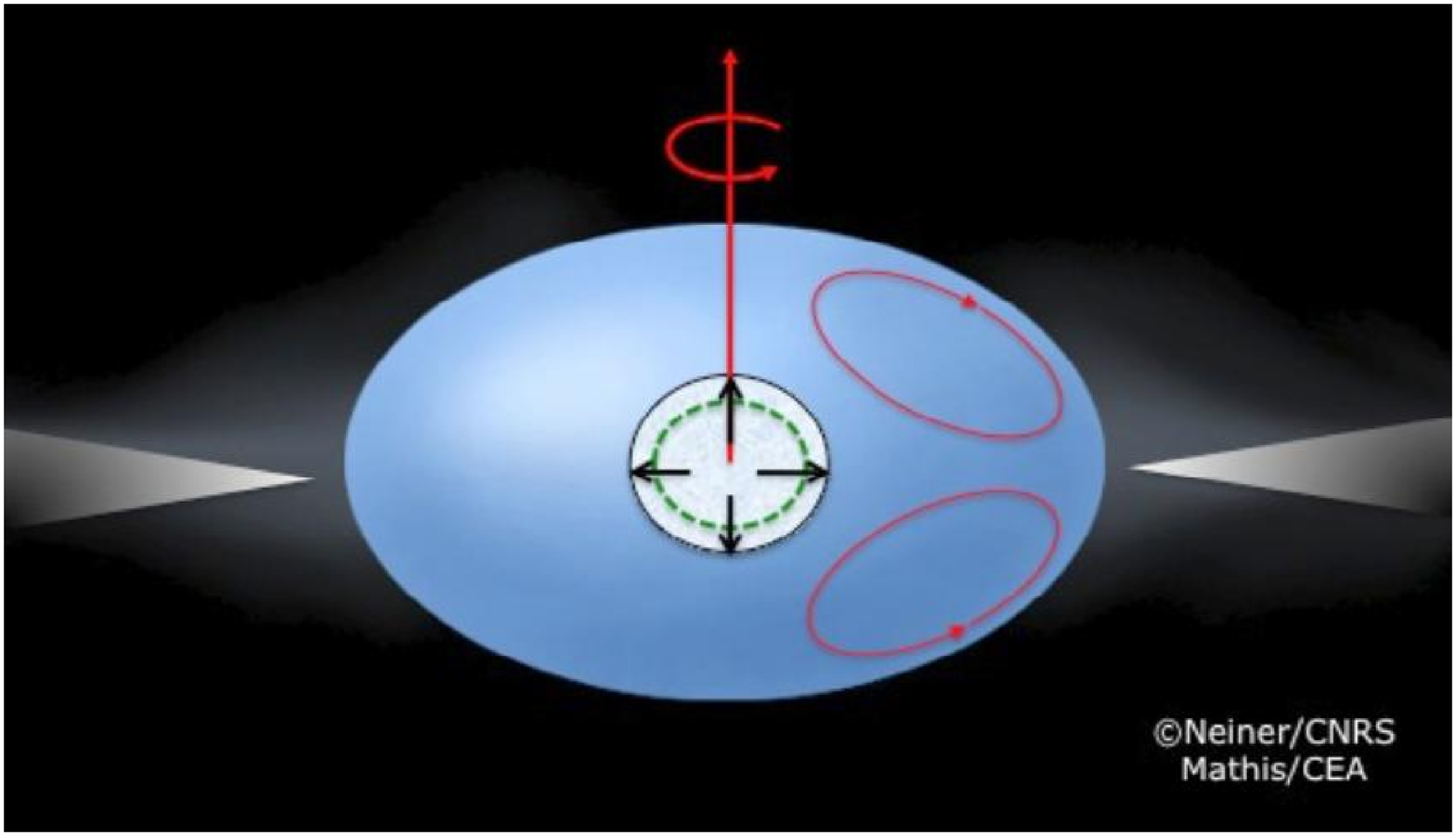}
\includegraphics[width=0.24\textwidth]{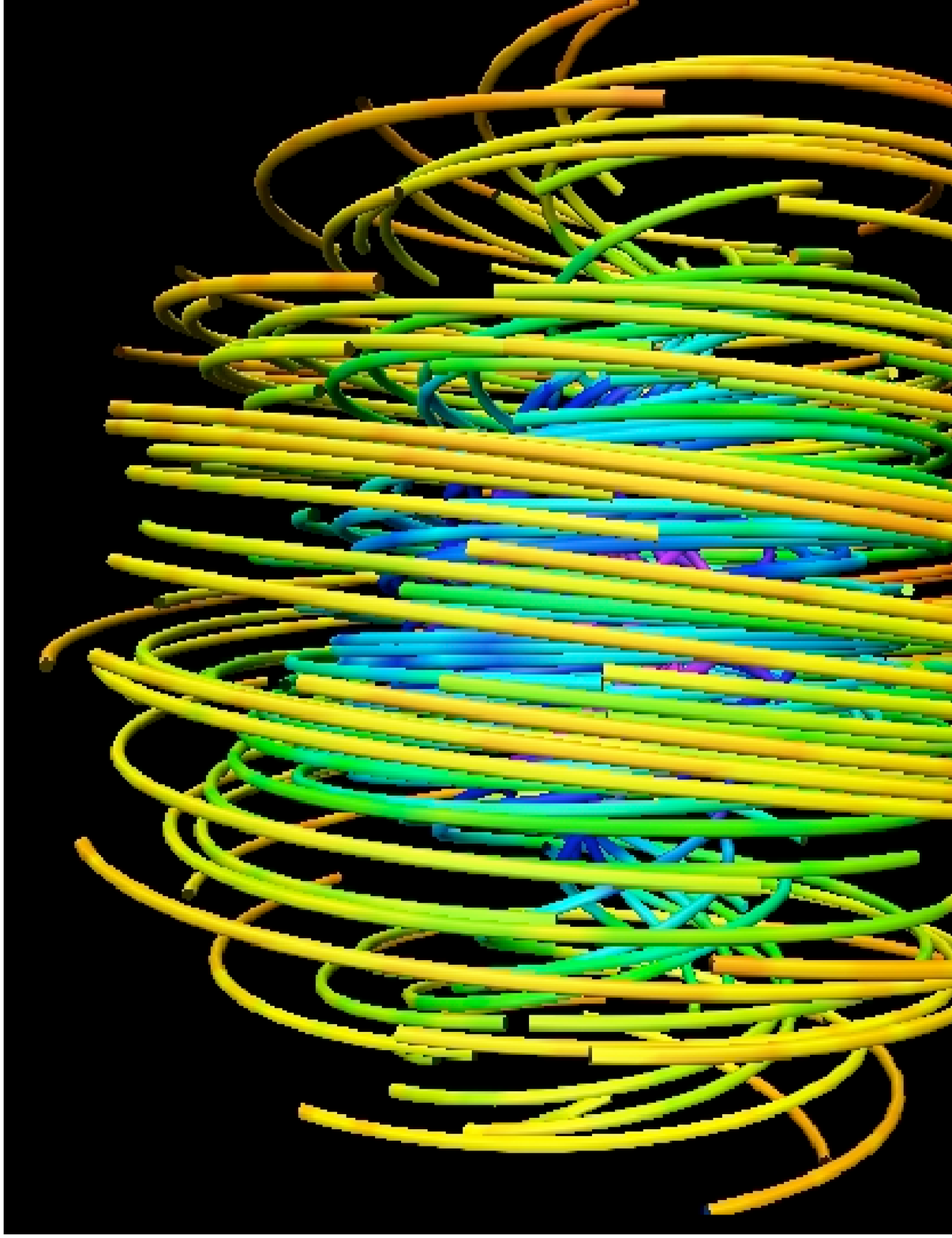} 
\caption{{\bf Left:} non-standard mixing processes in rapidly rotating massive stars: convective overshoot (black arrows) and rotational mixing (the red loops represent the large-scale meridional circulation driven by the differential rotation). The green dashed line corresponds to the "standard" limit of the convective core. {\bf Right}: Mixed stable fossil field configuration in a stellar radiation zone \citep[Taken from][]{DBM2010}.}
\label{fig1}
\end{figure}

HD\,181231 and HD\,175869 are two late rapidly rotating Be stars, which have been observed using high-precision photometry with the CoRoT satellite during about five consecutive months and 27 consecutive days, respectively. An analysis of their light curves, by \cite{Neineretal2009} and \cite{Gutetal2009} respectively, showed that several independent pulsation g-modes are present in these stars. Fundamental parameters have also been determined by these authors using spectroscopy. In \cite{Neineretal2012}, we succeeded to model these results to infer seismic properties of HD\,181231 and HD\,175869, and constrain internal transport and mixing processes of rapidly rotating massive stars. We used the non-adiabatic Tohoku oscillation code that accounts for the combined action of Coriolis and centrifugal accelerations on stellar pulsations as needed for rapid rotator modelling \citep[][]{LeeBaraffe1995}. The action of "non-standard" mixing processes was parametrized with the mixing parameter $\alpha_{\rm ov}$, which represents the "non-standard" extension of the convective core, and is determined by matching observed pulsation frequencies assuming a single star evolution scenario. We find that extra mixing of $\alpha_{\rm ov}= 0.3-0.35{\rm H_p}$, where ${\rm H_p}$ is the pressure scale-height, is needed in HD\,81231 and HD\,175869 to match the observed frequencies with those of prograde sectoral g-modes. We also detect the possible presence of r-modes. We investigated the respective contributions of several transport processes to this mixing. First, we used Geneva evolution models \citep{MM2000} to evaluate the contribution of the secular rotational transport and mixing processes in the radiative envelope, which is due to the combined action of differential rotation and of the shear-induced turbulence and meridional circulation it induces. Next, a Monte Carlo analysis of spectropolarimetric data was performed to examine the role of a potential fossil magnetic field. Finally, based on state-of-the-art modeling of penetrative convection at the top of the convective core and IGWs \citep{Browningetal2004}, we unravelled their respective contribution to the needed "non-standard" mixing. We showed that the extension of the convective core needed to match observations and models may be explained by mixing induced by the penetrative movements at the bottom of the radiative envelope ($\alpha_{\rm ov}= 0.2{\rm H_p}$) and by the secular hydrodynamical transport processes induced by the rotation in the envelope ($\alpha_{\rm ov}= 0.15{\rm H_p}$) (see fig. \ref{fig1}). This work showed how asteroseismology now opens a new door to probe transport and mixing processes in massive stars.

\subsection{Angular momentum in early-type stars as revealed by {\it Kepler}: a strong transport mechanism is needed}

The seismology of the Sun has allowed to probe its rotation profile until $0.2R_{\odot}$ and a quasiuniform rotation in the radiative core has been found, which strongly contrasts with the surrounding differentially rotating convective envelope. More recently, asteroseismology has allowed us to put constraints on rotation profiles in stars in the whole Hertzsprung-Russel diagram. In particular, \cite{Deheuvelsetal2012,Deheuvelsetal2014} and \cite{Becketal2012} have probed surface-to-core differential rotation in subgiant and red giant stars. Simultaneously, \cite{Mosseretal2012} has demonstrated the spin-down of the core of red giants
along their evolution. For upper-main sequence stars, \cite{Aertsetal2003}, \cite{Petal2004}, \cite{Betal2007}, \cite{DP2008}, \cite{Kurtzetal2014}, and \cite{Papicsetal2014} have provided constraints on the internal differential rotation. Finally, both rigid rotation \citep{Charpinetetal2009} and differential rotation \citep{Corsicoetal2011} have been discovered in white dwarfs. Therefore, as written by \cite{Kurtzetal2014}, these results "{\it have made angular momentum transport in stars throughout their life time an observational science}". This new transformational results all demonstrate that:
\begin{itemize}
\item i) Strong mechanism(s) for angular momentum transport must be acting along stellar evolution. Indeed, the simple conservation of angular momentum or purely hydrodynamical mechanisms related to differential rotation would lead to stronger angular velocity contrasts than those observed in stars \citep[e.g.][Eggenberger, this volume]{Ceillieretal2013}. Moreover their remnants (white dwarfs and neutron stars) would rotate faster than observed \citep[][]{Hegeretal2005,Suijsetal2008}.
\item ii) As demonstrated for example by the cases of the main-sequence A and SPB stars studied by \cite{Kurtzetal2014} and \cite{Papicsetal2014} respectively, it is absolutely necessary to go beyond models that treat the transport of angular momentum as purely diffusive processes as often done in the literature. Results have already been obtained theoretically by \cite{MM2000}, \cite{TC2005}, and \cite{MZ2005} for the transport of angular momentum by large-scale meridional flows, IGWs, and magnetic fields in stellar radiation zones, respectively. In the cases of the two stars studied by \cite{Kurtzetal2014} and \cite{Papicsetal2014}, the action of IGWs stochastically excited by the convective core (see Mathis \& Neiner, this volume) is a good candidate to explain the properties of observed rotation profiles \citep{Rogersetal2013,LNT2014}.
\end{itemize}

\section{Spectropolarimetry in upper main-sequence stars: a guide to understand fossil magnetism and its impact in stellar interiors}

Magnetic fields are now detected at the surface of $\sim$7\% of main-sequence and Pre-Main-Sequence intermediate-mass and massive stars \citep[][Grunhut, this volume]{Wadeetal2013,Alecianetal2013}. Indeed, fields ($300$ G to $30$ kG) are observed in some fraction of Herbig stars, A stars (the Ap stars), as well as in B stars and in a handful of O stars. Furthermore, non-convective neutron stars display fields strength of $10^8-10^{15}$ G. These magnetic fields in stably stratified non-convective stellar regions deeply modify the evolution of massive stars since their formation to their late stages. Their large-scale, stable, and ordered nature (often approximately dipolar) and the non-correlation of their properties with those of their host stars favour a fossil hypothesis (even if a dynamo is present in the convective core), whose origin has to be investigated. The first important point is to understand the topology of these large-scale magnetic fields. To survive since the star's formation or the PMS stage, they must be stable. It was suggested by \cite{Tayler1980} that a stellar magnetic field in stable axisymmetric equilibrium must contain both meridional and azimuthal components, since both are unstable on their own \citep{Tayler1973,MT1973}. This was confirmed by numerical simulations and theoretical works \citep[see fig. \ref{fig1} and ][Emeriau \& Mathis, this volume]{BN2006,DM2010} who showed that initial stochastic helical fields evolve on an Alfv\'en timescale into stable mixed configurations. This phenomenon well known in plasma physics is a MHD turbulent relaxation ({\it i.e.} a self-organization process involving magnetic reconnections in resistive MHD). After their formation, these fields deeply modify the transport of angular momentum and the mixing of chemicals. Indeed, they enforced a uniform rotation along field lines and the mixing is inhibited. This phenomena have been observed in the case of the star V2052 Oph studied using the combination of asteroseismology and spectropolarimetry \citep[][Neiner et al, this volume]{Briquetetal2012}.

\bibliographystyle{iau307}
\bibliography{Biblio_Mathis_proc4}

\begin{thebibliography}{}

\bibitem[\protect\astroncite{{Aerts} et~al.}{2010}]{Aertsetal2010}
{Aerts}, C., {Christensen-Dalsgaard}, J., \& {Kurtz}, D.~W. 2010,
\newblock {\em {Asteroseismology}}

\bibitem[\protect\astroncite{{Aerts} et~al.}{2003}]{Aertsetal2003}
{Aerts}, C., {Thoul}, A., {Daszy{\'n}ska}, J., {et~al.} 2003,
\newblock {\em Science} 300, 1926

\bibitem[\protect\astroncite{{Alecian} et~al.}{2013}]{Alecianetal2013}
{Alecian}, E., {Wade}, G.~A., {Catala}, C., {et~al.} 2013,
\newblock {\em \mnras} 429, 1001

\bibitem[\protect\astroncite{{Alvan} et~al.}{2013}]{AMD2013}
{Alvan}, L., {Mathis}, S., \& {Decressin}, T. 2013,
\newblock {\em \aap} 553, A86

\bibitem[\protect\astroncite{{Beck} et~al.}{2012}]{Becketal2012}
{Beck}, P.~G., {Montalban}, J., {Kallinger}, T., {et~al.} 2012,
\newblock {\em \nat} 481, 55

\bibitem[\protect\astroncite{{Braithwaite} \& {Nordlund}}{2006}]{BN2006}
{Braithwaite}, J. \& {Nordlund}, {\AA}. 2006,
\newblock {\em \aap} 450, 1077

\bibitem[\protect\astroncite{{Briquet} et~al.}{2007}]{Betal2007}
{Briquet}, M., {Morel}, T., {Thoul}, A., {et~al.} 2007,
\newblock {\em \mnras} 381, 1482

\bibitem[\protect\astroncite{{Briquet} et~al.}{2012}]{Briquetetal2012}
{Briquet}, M., {Neiner}, C., {Aerts}, C., {et~al.} 2012,
\newblock {\em \mnras} 427, 483

\bibitem[\protect\astroncite{{Browning} et~al.}{2004}]{Browningetal2004}
{Browning}, M.~K., {Brun}, A.~S., \& {Toomre}, J. 2004,
\newblock {\em \apj} 601, 512

\bibitem[\protect\astroncite{{Brun} et~al.}{2005}]{BBT2005}
{Brun}, A.~S., {Browning}, M.~K., \& {Toomre}, J. 2005,
\newblock {\em \apj} 629, 461

\bibitem[\protect\astroncite{{Ceillier} et~al.}{2013}]{Ceillieretal2013}
{Ceillier}, T., {Eggenberger}, P., {Garc{\'{\i}}a}, R.~A., \& {Mathis}, S.
  2013,
\newblock {\em \aap} 555, A54

\bibitem[\protect\astroncite{{Charpinet} et~al.}{2009}]{Charpinetetal2009}
{Charpinet}, S., {Fontaine}, G., \& {Brassard}, P. 2009,
\newblock {\em \nat} 461, 501

\bibitem[\protect\astroncite{{C{\'o}rsico} et~al.}{2011}]{Corsicoetal2011}
{C{\'o}rsico}, A.~H., {Althaus}, L.~G., {Kawaler}, S.~D., {et~al.} 2011,
\newblock {\em \mnras} 418, 2519

\bibitem[\protect\astroncite{{Decressin} et~al.}{2009}]{Detal2009}
{Decressin}, T., {Mathis}, S., {Palacios}, A., {et~al.} 2009,
\newblock {\em \aap} 495, 271

\bibitem[\protect\astroncite{{Deheuvels} et~al.}{2014}]{Deheuvelsetal2014}
{Deheuvels}, S., {Do{\u g}an}, G., {Goupil}, M.~J., {et~al.} 2014,
\newblock {\em \aap} 564, A27

\bibitem[\protect\astroncite{{Deheuvels} et~al.}{2012}]{Deheuvelsetal2012}
{Deheuvels}, S., {Garc{\'{\i}}a}, R.~A., {Chaplin}, W.~J., {et~al.} 2012,
\newblock {\em \apj} 756, 19

\bibitem[\protect\astroncite{{Donati} \&
  {Landstreet}}{2009}]{DonatiLandstreet2009}
{Donati}, J.-F. \& {Landstreet}, J.~D. 2009,
\newblock {\em \araa} 47, 333

\bibitem[\protect\astroncite{{Duez} et~al.}{2010}]{DBM2010}
{Duez}, V., {Braithwaite}, J., \& {Mathis}, S. 2010,
\newblock {\em \apjl} 724, L34

\bibitem[\protect\astroncite{{Duez} \& {Mathis}}{2010}]{DM2010}
{Duez}, V. \& {Mathis}, S. 2010,
\newblock {\em \aap} 517, A58

\bibitem[\protect\astroncite{{Dziembowski} \& {Pamyatnykh}}{2008}]{DP2008}
{Dziembowski}, W.~A. \& {Pamyatnykh}, A.~A. 2008,
\newblock {\em \mnras} 385, 2061

\bibitem[\protect\astroncite{{Featherstone} et~al.}{2009}]{Fetal2009}
{Featherstone}, N.~A., {Browning}, M.~K., {Brun}, A.~S., \& {Toomre}, J. 2009,
\newblock {\em \apj} 705, 1000

\bibitem[\protect\astroncite{{Garc{\'{\i}}a} et~al.}{2007}]{Garciaetal2007}
{Garc{\'{\i}}a}, R.~A., {Turck-Chi{\`e}ze}, S., {Jim{\'e}nez-Reyes}, S.~J.,
  {et~al.} 2007,
\newblock {\em Science} 316, 1591

\bibitem[\protect\astroncite{{Guti{\'e}rrez-Soto} et~al.}{2009}]{Gutetal2009}
{Guti{\'e}rrez-Soto}, J., {Floquet}, M., {Samadi}, R., {et~al.} 2009,
\newblock {\em \aap} 506, 133

\bibitem[\protect\astroncite{{Heger} et~al.}{2005}]{Hegeretal2005}
{Heger}, A., {Woosley}, S.~E., \& {Spruit}, H.~C. 2005,
\newblock {\em \apj} 626, 350

\bibitem[\protect\astroncite{{Kurtz} et~al.}{2014}]{Kurtzetal2014}
{Kurtz}, D.~W., {Saio}, H., {Takata}, M., {et~al.} 2014,
\newblock {\em ArXiv e-prints}

\bibitem[\protect\astroncite{{Lee} \& {Baraffe}}{1995}]{LeeBaraffe1995}
{Lee}, U. \& {Baraffe}, I. 1995,
\newblock {\em \aap} 301, 419

\bibitem[\protect\astroncite{{Lee} et~al.}{2014}]{LNT2014}
{Lee}, U., {Neiner}, C., \& {Mathis}, S. 2014,
\newblock {\em ArXiv e-prints}

\bibitem[\protect\astroncite{{Maeder}}{2009}]{Maeder2009}
{Maeder}, A. 2009,
\newblock {\em {Physics, Formation and Evolution of Rotating Stars}}

\bibitem[\protect\astroncite{{Markey} \& {Tayler}}{1973}]{MT1973}
{Markey}, P. \& {Tayler}, R.~J. 1973,
\newblock {\em \mnras} 163, 77

\bibitem[\protect\astroncite{{Mathis}}{2013}]{Mathis2013}
{Mathis}, S. 2013,
\newblock in M. {Goupil}, K. {Belkacem}, C. {Neiner}, F. {Ligni{\`e}res}, \&
  J.~J. {Green} (eds.), {\em Lecture Notes in Physics, Berlin Springer Verlag},
  Vol. 865 of {\em Lecture Notes in Physics, Berlin Springer Verlag}, p.~23

\bibitem[\protect\astroncite{{Mathis} \& {de Brye}}{2012}]{MdB2012}
{Mathis}, S. \& {de Brye}, N. 2012,
\newblock {\em \aap} 540, A37

\bibitem[\protect\astroncite{{Mathis} et~al.}{2013}]{Metal2013}
{Mathis}, S., {Decressin}, T., {Eggenberger}, P., \& {Charbonnel}, C. 2013,
\newblock {\em \aap} 558, A11

\bibitem[\protect\astroncite{{Mathis} et~al.}{2004}]{Metal2004}
{Mathis}, S., {Palacios}, A., \& {Zahn}, J.-P. 2004,
\newblock {\em \aap} 425, 243

\bibitem[\protect\astroncite{{Mathis} \& {Zahn}}{2004}]{MZ2004}
{Mathis}, S. \& {Zahn}, J.-P. 2004,
\newblock {\em \aap} 425, 229

\bibitem[\protect\astroncite{{Mathis} \& {Zahn}}{2005}]{MZ2005}
{Mathis}, S. \& {Zahn}, J.-P. 2005,
\newblock {\em \aap} 440, 653

\bibitem[\protect\astroncite{{Meynet} \& {Maeder}}{2000}]{MM2000}
{Meynet}, G. \& {Maeder}, A. 2000,
\newblock {\em \aap} 361, 101

\bibitem[\protect\astroncite{{Mosser} et~al.}{2012}]{Mosseretal2012}
{Mosser}, B., {Goupil}, M.~J., {Belkacem}, K., {et~al.} 2012,
\newblock {\em \aap} 548, A10

\bibitem[\protect\astroncite{{Neiner} et~al.}{2009}]{Neineretal2009}
{Neiner}, C., {Guti{\'e}rrez-Soto}, J., {Baudin}, F., {et~al.} 2009,
\newblock {\em \aap} 506, 143

\bibitem[\protect\astroncite{{Neiner} et~al.}{2012}]{Neineretal2012}
{Neiner}, C., {Mathis}, S., {Saio}, H., {et~al.} 2012,
\newblock {\em \aap} 539, A90

\bibitem[\protect\astroncite{{Pamyatnykh} et~al.}{2004}]{Petal2004}
{Pamyatnykh}, A.~A., {Handler}, G., \& {Dziembowski}, W.~A. 2004,
\newblock {\em \mnras} 350, 1022

\bibitem[\protect\astroncite{{P{\'a}pics} et~al.}{2014}]{Papicsetal2014}
{P{\'a}pics}, P.~I., {Moravveji}, E., {Aerts}, C., {et~al.} 2014,
\newblock {\em ArXiv e-prints}

\bibitem[\protect\astroncite{{Rogers} et~al.}{2013}]{Rogersetal2013}
{Rogers}, T.~M., {Lin}, D.~N.~C., {McElwaine}, J.~N., \& {Lau}, H.~H.~B. 2013,
\newblock {\em \apj} 772, 21

\bibitem[\protect\astroncite{{Suijs} et~al.}{2008}]{Suijsetal2008}
{Suijs}, M.~P.~L., {Langer}, N., {Poelarends}, A.-J., {et~al.} 2008,
\newblock {\em \aap} 481, L87

\bibitem[\protect\astroncite{{Talon} \& {Charbonnel}}{2005}]{TC2005}
{Talon}, S. \& {Charbonnel}, C. 2005,
\newblock {\em \aap} 440, 981

\bibitem[\protect\astroncite{{Tayler}}{1973}]{Tayler1973}
{Tayler}, R.~J. 1973,
\newblock {\em \mnras} 161, 365

\bibitem[\protect\astroncite{{Tayler}}{1980}]{Tayler1980}
{Tayler}, R.~J. 1980,
\newblock {\em \mnras} 191, 151

\bibitem[\protect\astroncite{{Wade} et~al.}{2013}]{Wadeetal2013}
{Wade}, G.~A., {Grunhut}, J., {Alecian}, E., {et~al.} 2013,
\newblock {\em ArXiv e-prints}

\bibitem[\protect\astroncite{{Zahn}}{1983}]{Zahn1983}
{Zahn}, J.-P. 1983,
\newblock in A.~N. {Cox}, S. {Vauclair}, \& J.~P. {Zahn} (eds.), {\em Saas-Fee
  Advanced Course 13: Astrophysical Processes in Upper Main Sequence Stars}, p.
  253

\bibitem[\protect\astroncite{{Zahn}}{1992}]{Zahn1992}
{Zahn}, J.-P. 1992,
\newblock {\em \aap} 265, 115

\bibitem[\protect\astroncite{{Zahn} et~al.}{2007}]{Zetal2007}
{Zahn}, J.-P., {Brun}, A.~S., \& {Mathis}, S. 2007,
\newblock {\em \aap} 474, 145

\end{thebibliography}

\end{document}